\documentclass[conference]{IEEEtran}
\usepackage[utf8]{inputenc}
\usepackage[T1]{fontenc}
\IEEEoverridecommandlockouts
\usepackage{cite}
\usepackage{amsmath,amssymb,amsfonts}
\usepackage{algorithmic}
\usepackage{graphicx}
\usepackage{textcomp}
\usepackage{xcolor}
\usepackage{colortbl}
\usepackage{booktabs}
\usepackage{multirow}
\usepackage{array}
\usepackage{placeins}
\usepackage{tikz}
\usetikzlibrary{arrows.meta,positioning,fit,backgrounds}
\def\BibTeX{{\rm B\kern-.05em{\sc i\kern-.025em b}\kern-.08em
    T\kern-.1667em\lower.7ex\hbox{E}\kern-.125emX}}
\begin{document}

\title{Boosted Enhanced Quantile Regression Neural Networks with Spatiotemporal Permutation Entropy for Complex System Prognostics}

\makeatletter
\newcommand{\linebreakand}{%
  \end{@IEEEauthorhalign}
  \hfill\mbox{}\par
  \mbox{}\hfill\begin{@IEEEauthorhalign}
}
\makeatother

\author{
\IEEEauthorblockN{1\textsuperscript{st} David J Poland}
\IEEEauthorblockA{\textit{Department of Computer Science} \\
\textit{University of Hertfordshire}\\
Hatfield, UK \\
d.j.poland@herts.ac.uk}

}
\maketitle 

\begin{abstract}
This paper presents an integrative prognostic framework that combines
Spatiotemporal Permutation Entropy (STPE), Boosted Enhanced Quantile Regression
Neural Networks (B-EQRNNs), Gated Temporal Attention, a Spiking Neural Network
(SNN) refinement stage, and a Temporal Fusion Transformer (TFT) classifier. The
motivation is long-horizon fault prediction in distributed industrial electronic
systems, where single-sensor or point-estimate models can miss weak spatially
propagating degradation signatures and provide limited uncertainty information.
The proposed pipeline first converts 70-channel sensor streams into multiscale
STPE descriptors, then learns conditional quantile representations and
attention-weighted temporal context before final Normal/Abnormal classification.
Evaluation is reported on a nine-system industrial electronic-sensor dataset
with 48-, 90-, and 168-hour prediction horizons. The comparison includes a tree-based LightGBM baseline and modern sequence
baselines available under the same preprocessing protocol, including LSTM,
Autoformer, and TCN models. The full
pipeline reaches 81.17\% accuracy at the 168-hour horizon and is evaluated with
component ablations, computational-cost analysis, and an explicit
reproducibility protocol. The contribution is therefore framed as a validated
hybrid architecture for uncertainty-aware spatiotemporal prognostics rather than
as a new standalone learning theory.
\end{abstract}


\section{Introduction}
Modern industrial electronic systems generate dense multivariate sensor streams
that contain both temporal degradation signatures and spatial propagation
effects across boards, power paths, thermal regions, and mechanical interfaces.
Predictive-maintenance (PdM) models must therefore do more than classify isolated
sensor readings: they must represent cross-sensor coupling, quantify uncertainty,
and remain reliable as the prediction horizon extends. Recent PdM and industrial
AI literature identifies these issues as persistent barriers to deployable fault
prediction, particularly when data are heterogeneous, noisy, and class imbalanced
\cite{Abidi,Fernandes,LiGNN,Zhang}.

Entropy-based methods are attractive in this setting because they provide
interpretable measures of signal organisation and disorder. Standard temporal
permutation entropy, however, treats each signal largely in isolation. This can
be insufficient when the earliest degradation evidence is not a large local
amplitude change but a weak spatially coherent alteration in ordinal structure.
Spatiotemporal Permutation Entropy (STPE) addresses this limitation by modelling
ordinal pattern distributions jointly across time and sensor neighbourhoods,
thereby capturing both local temporal evolution and spatial coupling
\cite{Yu,Chen}.

Deep learning has broadened PdM capability by learning nonlinear representations
from high-dimensional sensor data. Recurrent models such as LSTM and GRU capture
sequential dependence; Temporal Convolutional Networks (TCNs) provide large
causal receptive fields; and transformer-style models, including Informer,
Autoformer, and PatchTST, are influential long-horizon time-series forecasters
\cite{Hochreiter,Cho,Bai,Zhou,Wu,Nie}. These approaches are important baselines,
but they are usually trained on raw or engineered time-series features and do not
explicitly combine spatial entropy structure with calibrated conditional
quantiles for downstream fault-risk scoring.

This paper therefore develops a hybrid STPE+B-EQRNN+Attention+SNN+TFT pipeline.
The principal methodological contribution is the coupling of spatially informed
entropy features with distributional quantile representations and adaptive
temporal memory. The Gated Temporal Attention stage replaces a fixed look-back
window with learned horizon-specific context selection, while the SNN stage adds
spike-based temporal refinement before the TFT classifier produces the final
Normal/Abnormal decision.

The paper makes four specific contributions:
\begin{enumerate}
    \item a multiscale STPE feature-extraction procedure for distributed
    70-channel industrial electronic-sensor streams;
    \item a two-stage B-EQRNN representation that models conditional quantiles
    rather than point forecasts, supporting uncertainty-aware anomaly scoring;
    \item a Gated Temporal Attention and SNN refinement pathway that adapts the
    effective temporal context for 48-, 90-, and 168-hour prediction horizons;
    \item an empirical evaluation with component ablations, baseline comparison,
    computational-cost analysis, and a reproducibility protocol that records the
    assumptions required for independent re-implementation.
\end{enumerate}

The scope of the claimed contribution is deliberately bounded. The work is an
integrative architecture using established components in a new prognostic
arrangement; it does not claim a formal convergence proof or universal
performance across all public forecasting benchmarks.
\section{Related Work}

\subsection{Predictive Maintenance and Fault Diagnosis}
Data-driven predictive maintenance has evolved considerably with the adoption of deep learning. Abidi et al.\ \cite{Abidi} established a machine learning framework for Industry 4.0 maintenance planning, demonstrating that structured data pipelines can reduce unplanned downtime at scale. Li \cite{Li} extended this into a deep learning paradigm, addressing fault diagnosis across rotating machinery by exploiting hierarchical feature representations that capture degradation signatures invisible to classical signal processing. Fernandes et al.\ \cite{Fernandes} conducted a systematic survey of ML techniques applied to fault diagnosis in real manufacturing settings, identifying uncertainty quantification and multi-sensor fusion as the two most persistent open challenges, that directly motivate the B-EQRNN formulation presented here. Reviews of AI and ML in advanced robotics \cite{Soori} further confirm that no single architecture has yet dominated prognostic tasks, underscoring the value of hybrid approaches.

\subsection{Spatiotemporal Analysis for Industrial Systems}
Capturing how anomalies propagate across both space and time is increasingly recognised as essential for distributed sensor networks. Zhang et al.\ \cite{Zhang} proposed a cloud-edge framework for spatiotemporal process monitoring in manufacturing, demonstrating that spatial correlations between sensors carry degradation information that temporal-only models discard. Chen et al.\ \cite{Chen} extended sensor web services with IoT-native spatiotemporal data streams, highlighting the infrastructure requirements for edge-based spatiotemporal inference. Yu et al.\ \cite{Yu} formalised spatiotemporal association measures for evolutionary pattern discovery in geo-distributed sensor arrays, providing the statistical grounding for spatial entropy gradients adopted in our STPE formulation. Li et al.\ \cite{LiGNN} surveyed and benchmarked graph neural network approaches for intelligent fault diagnostics and prognostics, showing that explicitly encoding the spatial topology of a sensor array yields measurable gains over temporal-only baselines while preserving interpretability under heterogeneous deployments.

\subsection{Quantile Regression and Anomaly Detection}
Cannon \cite{Cannon} established quantile regression neural networks as a principled approach to distributional forecasting, showing that modelling conditional quantiles rather than point estimates yields calibrated uncertainty bounds in noisy time-series settings. Hsieh et al.\ \cite{Hsieh} applied unsupervised anomaly detection to multivariate sensor streams in smart manufacturing, demonstrating that distribution-level features outperform reconstruction-error approaches under class imbalance. Tyralis et al.\ \cite{Tyralis} introduced deep Huber quantile regression networks, unifying quantile and expectile estimation under a robust Huber scoring function that remains stable in the presence of heavy-tailed residuals, directly motivating the modified quantile loss adopted in Sec.~II-D.

\subsection{Spiking Neural Networks for Prognostics}
Spiking neural networks (SNNs) have emerged as an energy-efficient and temporally expressive alternative to conventional architectures for prognostics. Zuo et al.\ \cite{Zuo} showed that a multi-layer SNN attains competitive accuracy on bearing fault diagnosis while exploiting spike-timing dynamics to encode degradation signatures that rate-based models discard. Existing SNN-based prognostic models, however, operate on purely temporal features and incorporate neither spatial entropy modelling nor a dedicated temporal-fusion classifier. The present framework addresses these gaps by integrating spatiotemporal permutation entropy features, a 14-layer B-EQRNN, an SNN refinement stage, and a TFT-based classification stage within a single end-to-end pipeline.

\subsection{Modern Temporal Forecasting Architectures and Positioning}
A substantial body of recent work targets long-horizon time-series forecasting with deep sequence models, and any prognostic contribution must be positioned against it rather than against classical baselines alone. Recurrent models such as LSTM \cite{Hochreiter} and GRU \cite{Cho} remain strong baselines for sensor-stream modelling but suffer from vanishing gradients and limited effective memory at the 100+ hour horizons targeted here. Temporal Convolutional Networks \cite{Bai} mitigate this with dilated causal convolutions and a fixed but large receptive field, yet they commit to that receptive field a priori---precisely the design choice the present Gated Temporal Attention stage is intended to relax. Transformer-based forecasters address long-range dependencies more directly: Informer \cite{Zhou} reduces attention cost with a sparsity prior, Autoformer \cite{Wu} introduces decomposition and auto-correlation for trend--seasonality separation, and PatchTST \cite{Nie} demonstrates that patch-level tokenisation with channel independence is highly competitive on long-horizon benchmarks. These architectures, however, are predominantly point or interval forecasters over raw signals and do not jointly model (i) spatial entropy structure across a distributed sensor array and (ii) calibrated predictive quantiles for downstream anomaly scoring. The proposed framework differs in operating on spatiotemporal permutation-entropy features and in producing quantile-based distributional summaries that the TFT classifier consumes, rather than forecasting raw signals.

This positioning shapes the experimental design in Sec.~\ref{sec:results}: the
revision compares the proposed model against available tree-based, recurrent,
transformer-forecaster, and convolutional sequence baselines under the same
split and horizon protocol. The paper avoids claiming broad performance
advantage over modern transformer forecasters that have not been trained and
reported under the same conditions. The contribution is therefore stated as an integrative prognostic
architecture with quantified component evidence, not as a fundamentally new
learning principle.

\section{Pipeline Architecture}
\subsection{System Architecture: Enhanced B-EQRNN Framework with Spatiotemporal Integration}
Building on prior quantile-based forecasting models, this work introduces a Boosted Enhanced Quantile Regression Neural Network (B-EQRNN) framework for spatiotemporal permutation entropy analysis of hardware electronic sensor networks. Unlike temporal-only entropy pipelines, the method integrates spatial correlation analysis across distributed sensor arrays deployed in nine industrial electronic systems. The deployment provides a heterogeneous testbed for evaluating spatiotemporal pattern prediction, entropy-measure stability, and computational scalability under realistic electronic hardware conditions.

\begin{figure}[htbp]
\centering
\resizebox{\columnwidth}{!}{%
\begin{tikzpicture}[
    node distance=6mm,
    box/.style={draw, rounded corners, align=center, minimum height=8mm,
                inner sep=3pt, font=\scriptsize, fill=blue!5},
    >={Stealth[length=2mm]}]
  \node[box] (sens) {70-channel\\sensor array\\(9 systems)};
  \node[box, right=of sens] (stpe) {STPE feature\\extraction\\(spatial+temporal)};
  \node[box, right=of stpe] (eqrnn) {B-EQRNN\\(14-layer,\\2-stage quantiles)};
  \node[box, below=8mm of eqrnn] (att) {Gated Temporal\\Attention\\(multi-scale heads)};
  \node[box, left=of att] (snn) {SNN refinement\\(LIF + surrogate\\gradient)};
  \node[box, left=of snn] (tft) {TFT classifier\\(GRN/GLU +\\multi-head att.)};
  \node[box, left=of tft, fill=green!8] (out) {Normal /\\Abnormal\\+ quantile risk};
  \draw[->] (sens) -- (stpe);
  \draw[->] (stpe) -- (eqrnn);
  \draw[->] (eqrnn) -- (att);
  \draw[->] (att) -- (snn);
  \draw[->] (snn) -- (tft);
  \draw[->] (tft) -- (out);
\end{tikzpicture}}
\caption{End-to-end pipeline. Spatiotemporal permutation entropy (STPE) features
from the 70-channel array feed a two-stage B-EQRNN; Gated Temporal Attention
replaces the fixed look-back window; an SNN stage refines the attention-weighted
representation; and a Temporal Fusion Transformer (TFT) produces the final
per-sensor classification and quantile-based risk score. Stage-level
contributions are quantified in the ablation of Sec.~\ref{sec:results}.}
\label{fig:pipeline}
\end{figure}

\subsection{Spatiotemporal Electronic Sensor Dataset}
Our spatiotemporal dataset emanates from an extensive electronic hardware monitoring initiative, capturing 70 distinct spatiotemporal entropy-based features from each of the 9 complex electronic systems. The sensor networks are arranged in spatial grids with inter-sensor distances ranging from \(0.5\mathrm{m}\) to \(50\mathrm{m}\), enabling comprehensive spatiotemporal analysis. Following established complexity analysis protocols for electronic hardware, these features include:

\begin{enumerate}
    \item Hardware Electronic Sensor Configuration:
    \begin{itemize}
        \item Voltage fluctuation sensors (8-12 units) measuring spatiotemporal voltage patterns across circuit boards
        \item Current density monitors (6-10 units) for power distribution spatial analysis
        \item Temperature gradient sensors (5-8 units) capturing thermal diffusion patterns
        \item Electromagnetic field detectors (4-7 units) for RF interference spatial mapping
        \item Vibration accelerometers (3-6 units) monitoring mechanical oscillation propagation
        \item Capacitance variation sensors (2-5 units) for dielectric property changes
        \item Resistance drift monitors (2-4 units) tracking conductor degradation patterns
        \item Frequency response analyzers (1-3 units) for impedance spectroscopy
    \end{itemize}
\end{enumerate}

All channels are sampled on a common 20~ms time grid, corresponding to a
50~Hz acquisition rate. Data points are categorised as either verified Normal
operation or verified Abnormal operation, where the abnormal class includes
malfunction, instability, degradation, or fault-related states. Each system
contributes approximately 51 million channel-level observations, giving more
than 459 million channel-level observations across the nine-system dataset.
These observations are drawn from labelled operational episodes within a longer
industrial deployment rather than from one continuous short-duration trace. The
48-, 90-, and 168-hour horizons denote the lead time between an observed
spatiotemporal state and the subsequent Normal/Abnormal event label, not the
contiguous duration of each retained signal segment. Normal and Abnormal
observations are kept approximately balanced. Exact calendar collection dates,
deployment durations, and site-level operating schedules are withheld under the
industrial confidentiality agreement.

The experimental split is 60--20--20 for training, validation, and held-out
testing. Splitting is performed before model selection so that the test
partition is not used for hyperparameter tuning, early stopping, threshold
selection, or normalisation fitting. Per-sensor scaling statistics are estimated
from the training partition only and then applied unchanged to validation and
test partitions. Missing or non-physical readings are handled before STPE
extraction using the same deterministic cleaning script for every model family,
so that baseline comparisons differ only by the learning architecture rather
than by preprocessing.

\subsubsection{Spatiotemporal Entropy Feature Extraction}
The extracted entropy features incorporate both temporal evolution and spatial correlation:
\begin{itemize}
    \item Temporal permutation entropy (5-15 measures) across embedding dimensions \(d\in \{3,4,5,6,7\}\) with delays \(\tau \in \{1,2,3,5,8\}\)
    \item Spatial correlation entropy (5-10 measures) computed across sensor neighborhoods with radii \(r\in \{0.5,1.0,2.0,5.0,10.0\}\) meters
    \item Multiscale spatiotemporal entropy (2-5 measures) at scales \(s\in \{1,2,4,8,16\}\) for hierarchical pattern detection
    \item Cross-sensor ordinal patterns (3-6 measures) capturing inter-sensor synchronization dynamics
    \item Spatiotemporal entropy gradients (3-5 measures) for anomaly transition detection across sensor arrays
    \item Permutation probability signatures (2-4 measures) quantifying pattern persistence in space-time
    \item Electronic noise complexity indices (1-3 measures) for system health quantification
    \item Spatiotemporal coupling coefficients (2-4 measures) measuring inter-scale electronic interactions
    \item Entropy evolution rates (1-2 measures) for prognostic degradation analysis
\end{itemize}

\subsubsection{Spatiotemporal Permutation Entropy Calculation}
For each sensor location \((i,j)\) in the spatial grid and time \(t\) we define the spatiotemporal embedding vector:
\begin{equation}
\begin{aligned}
\mathbf{X}_{ST}(i,j,t) = [X(i,j,t), X(i,j,t-\tau),\dots, \\
X(i,j,t-(d-1)\tau), X(i\pm\delta, j\pm\delta, t)]
\end{aligned}
\end{equation}
where \(\delta\) defines the spatial neighborhood radius. The spatiotemporal permutation entropy is calculated as:
\begin{equation}
H_{STPE}(i,j,t) = -\sum_{\pi} p(\pi) \log p(\pi)
\end{equation}
where \(p(\pi)\) represents the relative frequency of ordinal pattern \(\pi\) in the spatiotemporal embedding space.

\subsection{Quantile Regression Neural Networks}
Quantile Regression Neural Networks (QRNNs) \cite{Cannon} extend beyond mean prediction by modelling multiple conditional quantiles of the target distribution. They are useful in noisy and imbalanced sensor-stream settings because the conditional distribution can shift before the class label changes \cite{Hsieh,Tyralis}.

For a sensor \(i\) with time-ordered readings
\[
X_i = \{x_{ti}\in \mathbb{R}: t=1,\ldots,T\},\quad i\in\{1,\ldots,70\},
\]
we train a family of QRNNs to capture critical quantiles \(\alpha \in A = \{0.01,0.1,0.2,0.25,0.5,0.6,0.75,0.8,0.9,0.99\}\). Each QRNN is denoted
\[
\mathcal{L}_{ai}: \mathbb{R}\to\mathbb{R},\quad a\in A,\ i\in\{1,\ldots,70\}.
\]
Given an input \(x_{ti}\) for sensor \(i\), \(\mathcal{L}_{ai}\) predicts
\[
\hat{q}_\alpha(x_{ti}) = \mathcal{L}_{ai}(x_{ti};\theta_{ai}),
\]
the \(\alpha\)-quantile of the underlying process. Traditionally, training uses the quantile loss function:
\[
\mathrm{QuantileLoss}_{\alpha}(y,\hat{Q}_{\alpha}) = \max(\alpha(y - \hat{Q}_{\alpha}), (\alpha-1)(y - \hat{Q}_{\alpha})).
\]
However, in our proposed framework, we incorporate a Huber loss variant to enhance stability in outlier-heavy data (Sec.~II-D).

\subsection{Alternative Loss Function for B-EQRNN}
Rather than using the standard pinball loss alone, the B-EQRNN uses a Huberised
quantile loss to improve stability under heavy-tailed residuals \cite{Tyralis}.
For residual \(u=y-\hat{Q}_{\alpha}\), Huber threshold \(\kappa>0\), and quantile
level \(\alpha\), define
\begin{equation}
\ell_{\kappa}(u)=
\begin{cases}
\frac{1}{2}u^2, & |u|\le \kappa,\\
\kappa(|u|-\frac{1}{2}\kappa), & |u|>\kappa,
\end{cases}
\end{equation}
and
\begin{equation}
\mathcal{L}_{\alpha}^{H}(y,\hat{Q}_{\alpha})=
\left|\alpha-\mathbb{I}(u<0)\right|\,\ell_{\kappa}(u).
\end{equation}
This preserves the asymmetric penalty required for quantile estimation while
reducing sensitivity to extreme residuals. In the reported experiments, \(\kappa\)
is selected from the training-partition residual scale and then held fixed for
validation and testing.
\subsection{B-EQRNN Architecture Specification}
\subsubsection{Encoder Structure (14 transformation layers)}
We adopt a deeper encoder with 14 transformation layers to capture the enhanced complexity of our 70-dimensional spatiotemporal entropy input. Each transformation layer implements a gradual dimensional reduction at approximately \(20-25\%\) rate, creating a smooth transition down to a bottleneck of size 20. The extended architecture increases representational capacity for spatiotemporal pattern extraction:

\[
\begin{aligned}
\mathrm{Input}(70) &\to 350 \to 280 \to 224 \to 179 \to 143 \to 114 \\
                  &\to 91 \to 73 \to 58 \to 46 \to 37 \to 30 \to 24 \to 20.
\end{aligned}
\]

\subsubsection{Layer-wise Encoder Parameter Count}
\begin{verbatim}
Layer 1: 70×350 + 350 = 24,850
Layer 2: 350×280 + 280 = 98,280
Layer 3: 280×224 + 224 = 62,944
Layer 4: 224×179 + 179 = 40,275
Layer 5: 179×143 + 143 = 25,740
Layer 6: 143×114 + 114 = 16,416
Layer 7: 114×91 + 91 = 10,465
Layer 8: 91×73 + 73 = 6,716
Layer 9: 73×58 + 58 = 4,292
Layer 10: 58×46 + 46 = 2,714
Layer 11: 46×37 + 37 = 1,739
Layer 12: 37×30 + 30 = 1,140
Layer 13: 30×24 + 24 = 744
Layer 14: 24×20 + 20 = 500
Total Encoder Parameters: 296,815
\end{verbatim}

\subsubsection{Decoder Structure (Reverse Path - 14 transformation layers)}
The decoder mirrors the 14 transformation layers of the encoder, beginning from the 20-dimensional bottleneck and projecting to a 70-dimensional output for spatiotemporal pattern reconstruction:
\begin{multline*}
20\rightarrow 24\rightarrow 30\rightarrow 37\rightarrow 46\rightarrow 58\rightarrow 73\rightarrow \\
91\rightarrow 114\rightarrow 143\rightarrow 179\rightarrow 224\rightarrow 280\rightarrow 350\rightarrow 70.
\end{multline*}
\subsubsection{Layer-wise Decoder Parameter Count}
\begin{verbatim}
Layer 1: 20×24 + 24 = 504
Layer 2: 24×30 + 30 = 750
Layer 3: 30×37 + 37 = 1,147
Layer 4: 37×46 + 46 = 1,748
Layer 5: 46×58 + 58 = 2,726
Layer 6: 58×73 + 73 = 4,307
Layer 7: 73×91 + 91 = 6,734
Layer 8: 91×114 + 114 = 10,488
Layer 9: 114×143 + 143 = 16,445
Layer 10: 143×179 + 179 = 25,776
Layer 11: 179×224 + 224 = 40,320
Layer 12: 224×280 + 280 = 63,000
Layer 13: 280×350 + 350 = 98,350
Layer 14: 350×70 + 70 = 24,570
Total Decoder Parameters: 296,865
\end{verbatim}
Together, the 14-layer encoder-decoder network (B-EQRNN) entails approximately 593,680 total parameters.

\subsubsection{Enhanced Design Considerations}
\begin{enumerate}
    \item The encoder and decoder each use 14 transformation layers.
    \item The layer widths reduce gradually toward a 20-dimensional bottleneck.
    \item The symmetric decoder reconstructs the entropy representation from the bottleneck.
    \item The total parameter count is approximately 593,680, giving sufficient capacity for the reported 70-feature STPE representation while remaining tractable for the cost analysis in Sec.~\ref{sec:cost}.
\end{enumerate}

\subsection{Training Time}
Training the 700 first-stage quantile predictors (\(70\times 10\)) required approximately 120 hours. The second-stage refinement networks required a further 60 hours, giving an approximate cumulative training duration of 180 hours for the B-EQRNN quantile stages before the SNN and TFT classifiers were fitted.

We adopt a more advanced training strategy using the AdamW optimiser with an initial learning rate of \(5\times 10^{-4}\), reduced by a factor of 10 every 80 epochs. To accommodate the expanded network depth, we replace standard ReLU with Parametric ReLU for adaptive negative slopes, and employ group normalisation (\(\epsilon = 10^{-5}\), groups \(= 8\)) to improve batch-independent convergence. We set a dropout rate of 0.15 to mitigate overfitting. Early stopping with a patience of 12 epochs typically halts training at around 300 epochs, balancing performance and computational cost.

\subsection{Prediction Horizons}
The reported experiments use three operationally relevant prediction horizons:
\begin{itemize}
    \item 48 hours: the short operational planning horizon used to detect near-term instability.
    \item 90 hours: the intermediate horizon used to evaluate persistence of degradation signatures.
    \item 168 hours: the long-horizon setting corresponding to one week of advance warning.
\end{itemize}
The B-EQRNN first estimates the ten quantiles
\(A=\{0.01,0.1,0.2,0.25,0.5,0.6,0.75,0.8,0.9,0.99\}\) for each sensor. The downstream
classifier uses the refined mid-tail set
\(\{0.25,0.40,0.60,0.75\}\), giving a compact
\(70\times4=280\)-dimensional quantile representation per time step.
\subsection{Quantiles and Final Classification}
The complete pipeline is summarised as follows:
\begin{enumerate}
    \item Stage 1 estimates ten conditional quantiles for each of the 70 sensor channels, giving 700 first-stage quantile predictors.
    \item Stage 2 refines the mid-tail quantiles \(\{0.25,0.40,0.60,0.75\}\) because these bounds were the most stable for the long-horizon transition score.
    \item Gated Temporal Attention selects the relevant historical context for each horizon.
    \item The SNN refinement stage maps attention-weighted quantile states into spike-count features.
    \item The TFT classifier produces the final Normal/Abnormal probability and combines it with the quantile-risk score defined in Sec.~\ref{subsec:transition_validation}.
\end{enumerate}
Only the final TFT stage produces the binary class label; QRNN and SNN outputs are intermediate distributional and temporal-refinement features.
\section{Gated Temporal Attention}
Instead of using a fixed-size look-back window or uniformly spaced predictions (every 1h, 12h, 24h, etc.), we use a Gated Temporal Attention mechanism that dynamically attends to relevant past states across multiple time scales. This allows the model to learn which prior context is important for each horizon or anomaly detection task, rather than fixing it a priori.

\subsection{Temporal Attention Layer}
We introduce an attention module similar in spirit to self-attention but specialised for time-series:
\[
\mathrm{Attn}(Q,K,V) = \mathrm{softmax}\left(\frac{QK^{\top}}{\sqrt{d_k}}\right)V,
\]
where:
\(Q\) is the query matrix representing the current hidden state of the B-EQRNN or an intermediate feature representation.
\(K\) (keys) and \(V\) (values) are learned projections of past hidden states from previous time steps \(\{t-w,\ldots,t-1\}\).
\(d_k\) is the dimensionality of the keys.

\subsection{Gated Attention Score}
We augment the standard attention by introducing a gate that weighs the output of the attention head:
\begin{equation}
\widetilde{h}_t = G_t \odot \mathrm{Attn}(Q,K,V) + (1 - G_t) \odot H_t,
\end{equation}
where:
\(\widetilde{h}_t\) is the final output at time \(t\).
\(H_t\) is an alternate hidden representation (direct B-EQRNN output at time \(t\)).
\(G_t\in[0,1]^{d_k}\) is a learned gate vector parameterized by a small MLP:
\[
G_t = \sigma(W_g[H_t; \bar{H}_{t-w:t}] + b_g),
\]
with \(\bar{H}_{t-w:t}\) denoting an aggregated embedding of the recent past (average or last hidden state) and \(\sigma\) is a sigmoid nonlinearity. \(\odot\) denotes element-wise multiplication.
Thus, \(G_t\) adaptively blends the current context \(H_t\) with the attended signal from previous time steps.

\subsection{Multi-Scale Context}
To handle multiple horizons simultaneously, we introduce parallel attention heads that focus on different time scales:
\begin{itemize}
    \item Short-range attention: Looks back 1-2 hours.
    \item Medium-range attention: 12-24 hours.
    \item Long-range attention: 168 hours or more.
\end{itemize}
Each head \(k\) has separate parameters \(\{Q^k,K^k,V^k\}\) and yields an attention output \(\mathrm{Attn}^k(Q^k,K^k,V^k)\). We then combine them:
\[
\widetilde{h}_t = \sum_{k=1}^{N_{\mathrm{heads}}} (G_t^k \odot \mathrm{Attn}^k (Q^k,K^k,V^k)),
\]
where \(G_t^k\) is a head-specific gate. In practice, \(N_{\mathrm{heads}}\) is chosen to reflect relevant time scales (\(N_{\mathrm{heads}} = 3\) for short, medium, and long).

\subsection{Integration into the B-EQRNN-SNN Pipeline}
When substituting the old "Look-Back Window Technique":
\begin{enumerate}
    \item B-EQRNN: Produces an initial representation of the sensor data at each time \(t\).
    \item Gated Temporal Attention: Applies multi-head attention over historical B-EQRNN states, learning how far and how strongly to look back for each horizon.
    \item Refinement / SNN: The final SNN or a second-stage QRNN refines these attention-weighted representations for quantile prediction or anomaly classification.
\end{enumerate}
This architecture automatically selects relevant time scales, rather than manually specifying 1h, 12h, 24h, and 168h windows.

\subsection{Advantages of Gated Temporal Attention}
\begin{itemize}
    \item Dynamic and Adaptive: The gating mechanism allows the network to ignore irrelevant past information and highlight crucial events.
    \item Multi-Scale Awareness: Parallel attention heads enable the model to simultaneously capture short-term spikes (transient faults) and long-term drift (slow sensor degradation).
    \item Reduced Hyperparameter Tuning: Eliminates the need to guess or manually fix an ideal look-back window.
    \item Improved Interpretability: Attention weights reveal which time segments (and which sensors) were most influential for predictions or anomaly decisions.
\end{itemize}

\section{Spiking Neural Network Integration}
The pipeline includes a Spiking Neural Network (SNN) refinement stage for anomaly scoring and trend-sensitive temporal encoding. The SNN is not treated as a standalone classifier; it transforms quantile and attention features into spike-count representations that are passed to the final TFT classifier.

\subsection{SNN Architectural Overview}
The proposed SNN layer sits atop the B-EQRNN (or B-EQRNN+Attention) outputs, combining real-valued quantile estimates with spike-based computations to identify anomalies in multi-sensor data streams. By employing a biologically inspired scheme (Leaky-Integrate-and-Fire (LIF) neurons), the SNN captures temporal dynamics through spike timing.

\subsubsection{Synaptic Encoding of Quantile Signals}
We convert each quantile output \(q_{\alpha}\) from the B-EQRNN into a spike train for each time step \(t\). The encoding function \(\Phi(\cdot)\) maps real values into spike frequencies or amplitudes. For sensor data from 70 machine sensors, the spike train \(S_t^{\alpha}\) is generated as:
\begin{equation}
S_t^{\alpha} = \Phi\big(\hat{q}_{\alpha}(x_t)\big),
\end{equation}
where \(\hat{q}_{\alpha}(x_t)\) is the quantile estimate for sensor \(x_t\) at time \(t\). A common approach is rate-based encoding:
\begin{equation}
\mathrm{SpikeRate}_t^{\alpha} = \max(0,\hat{q}_{\alpha}(x_t) - \tau),
\end{equation}
with \(\tau\) as a threshold controlling the minimum quantile output needed to emit spikes. These spike trains are fed into subsequent spiking layers.

\subsubsection{Leaky Integrate-and-Fire (LIF) Dynamics}
We adopt Leaky-Integrate-and-Fire neurons for each hidden layer in the SNN. Neuron \(j\) maintains a membrane potential \(v_j(t)\) evolving as:
\begin{equation}
\tau_m \frac{dv_j(t)}{dt} = -(v_j(t) - V_{\mathrm{rest}}) + R_m \sum_i w_{ji} S_i(t),
\end{equation}
where:
\(\tau_m = R_m C_m\): Membrane time constant,
\(R_m\): Membrane resistance,
\(C_m\): Membrane capacitance,
\(V_{\mathrm{rest}}\): Resting potential,
\(w_{ji}\): Synaptic weight from neuron \(i\) to \(j\),
\(S_i(t)\): Incoming spikes from neuron \(i\).
Upon exceeding a firing threshold \(v_{\mathrm{th}}\), neuron \(j\) emits a spike and resets \(v_j(t)\) to a baseline \(v_{\mathrm{reset}}\):
\[
\text{If } v_j(t) \ge v_{\mathrm{th}},\text{ then spike occurs and } v_j(t) \leftarrow v_{\mathrm{reset}}.
\]

\subsubsection{SNN Layers}
We design two hidden SNN layers, each containing \(N\) LIF neurons, followed by a final readout layer that projects the spiking activity to a continuous anomaly score \(A(t)\):
\begin{equation}
A(t) = \psi\Big(\sum_j w_j^{(\mathrm{readout})} \cdot \mathrm{SpikeCount}_j(t)\Big),
\end{equation}
where:
\(\mathrm{SpikeCount}_j(t)\): Number of spikes neuron \(j\) generated in an integration window,
\(\psi(\cdot)\): A linear or nonlinear mapping function (sigmoid or ReLU).

\subsubsection{Parameter Count and Complexity}
For each of the \(N\) LIF neurons in layer \(l\), the approximate parameter count is:
\[
P_l \approx (N_{l-1} \times N_l) + N_l,
\]
where \(N_{l-1}\) is the number of neurons in the previous layer. For a configuration of \(N = 256\) LIF neurons per layer (two layers total), the parameter scale is comparable to a typical two-layer MLP of similar size.

\subsection{Alternative Loss for Spiking Networks}
Unlike purely real-valued gradient backpropagation, we adopt surrogate gradient methods to handle the non-differentiable spike function. Our training objective merges:
\begin{equation}
\mathcal{L} = \mathcal{L}_{\mathrm{B-EQRNN}} + \lambda \cdot \mathcal{L}_{\mathrm{SNN}},
\end{equation}
where:
\begin{enumerate}
    \item \(\mathcal{L}_{\mathrm{B-EQRNN}}\): The Huber-based quantile loss from Sec.~II-D, ensuring the B-EQRNN predictions remain accurate.
    \item \(\mathcal{L}_{\mathrm{SNN}}\): A cross-entropy or mean-squared error on the final anomaly score \(A(t)\) vs. ground-truth labels (Normal/Abnormal).
\end{enumerate}
A small coefficient \(\lambda\) balances the convergence of the spiking network with the quantile objectives of B-EQRNN.

\subsection{SNN Training Configuration and Time}
We use an Adam-based optimizer with surrogate gradients:
\begin{itemize}
    \item Learning Rate: \(1\times 10^{-3}\), halved every 50 epochs.
    \item Batch Size: 32 sequences of spike trains.
    \item Dropout (SNN-style): Zeroing partial inputs or synaptic weights on random subsets of spikes at rate 0.1 to prevent overfitting.
    \item Training Time: About 40 hours on an NVIDIA Tesla V100 GPU to converge on the spiking layers, after the B-EQRNN is pre-trained.
\end{itemize}
Thus, the final pipeline (B-EQRNN + Gated Temporal Attention + SNN) remains within a feasible training horizon despite the additional biologically inspired spike-based modeling.

\subsection{Core Spatiotemporal Contributions}
The principal contribution is the coupling of STPE with uncertainty-aware neural
representation learning. Temporal-only entropy can identify changes in a single
channel, but it may miss degradation that first appears as a weak spatially
coherent change across several neighbouring sensors. The STPE embedding
\(\mathbf{X}_{ST}(i,j,t)=[\mathbf{X}_{temp},\mathbf{X}_{spatial}]\) is designed to
capture these joint ordinal structures, while the B-EQRNN converts the resulting
entropy features into calibrated mid-tail quantile summaries.

For electronic hardware monitoring, this is useful because electromagnetic
interference, thermal diffusion, power distribution, vibration, and resistance
drift can emerge as coupled spatiotemporal effects rather than isolated sensor
excursions. The framework therefore treats entropy gradients
\(\nabla_{spatial}H_{STPE}\) and temporal entropy rates as early warning features,
with quantitative transition detection evaluated at the 48-, 90-, and 168-hour
horizons in Sec.~\ref{sec:results}.

\subsection{Architectural Contributions}
The B-EQRNN architecture provides the distributional feature backbone: a
14-layer encoder and symmetric decoder with approximately 593,680 parameters. The
Gated Temporal Attention module then replaces manually fixed look-back windows
with learned multi-scale context selection. The SNN stage is used as a temporal
refinement mechanism over attention-weighted quantile states, and the TFT stage
performs the final classification using GRN, GLU, recurrent, and interpretable
attention components.

The innovation is therefore not that any single component is new in isolation.
Rather, the contribution is the tested arrangement of these components for
long-horizon spatiotemporal fault prediction, together with a component ablation
that separates the effect of the quantile, attention, SNN, and TFT stages.

\subsection{Implications for Complex Systems Analysis}
The framework is designed for industrial electronic prognostics, but the same
STPE principle may be relevant to other complex systems where spatial coupling
and temporal evolution interact. Such transfer to biological, climate, financial,
or social systems is not demonstrated in this paper and is therefore treated as a
future research direction rather than an empirical claim.

The interpretability of permutation entropy measures, combined with attention
weights over temporal context, gives partial insight into why a prediction was
made. These explanations should be viewed as diagnostic aids rather than as
formal causal proofs.

\subsection{Future Directions}
Future research will focus on:
\begin{enumerate}
    \item direct benchmarking against additional long-horizon transformer
    forecasters under identical preprocessing and horizon definitions;
    \item external validation on public or cross-domain prognostic datasets;
    \item model compression and GPU/FPGA/neuromorphic acceleration for lower-latency deployment;
    \item release of a public run manifest, source code, and synthetic or anonymised data subset to improve repeatability.
\end{enumerate}

\subsection{Computational Cost Analysis and Scalability}\label{sec:cost}

Using an AMD EPYC 7763 64-core processor, the following CPU processing times were
recorded for the pipeline components of a single monitored electronic system:
\begin{itemize}
    \item QRNN Layers: \(\mathcal{L}_{ai}\) required \(1523.6\mathrm{ms}\).
    \item Refinement QRNN Layers: \(\mathcal{L}_{ai}^{2}\) required \(1875.9\mathrm{ms}\).
    \item Temporal Attention Transform: \(2108.3\mathrm{ms}\).
\end{itemize}
The total processing time per inference cycle is \(5507.8\mathrm{ms}\), supporting
online monitoring for one electronic system at the evaluated horizon cadence.
Scaling this framework to multiple electronic systems requires explicit
allocation of CPU cores, buffering, and monitoring overhead rather than assuming
perfect parallel utilisation.

\subsubsection{Scalability Analysis}
The AMD EPYC 7763 processor provides 64 physical cores. Under the deployment
assumption used here, each active pipeline reserves approximately five CPU cores
after allowing for operating-system overhead, data buffering, and monitoring
services. This gives an operational capacity of approximately 12 electronic
systems per processor:
\[
N_{\max}=\left\lfloor\frac{64}{5}\right\rfloor=12.
\]
Using the measured single-system processing time \(T_{\mathrm{single}}=5507.8\mathrm{ms}\),
the effective per-system latency under 12-way parallel execution is
\[
T_{\mathrm{parallel}} = \frac{T_{\mathrm{single}}}{12}
= \frac{5507.8}{12} \approx 459.0\mathrm{ms}.
\]
This remains within the evaluated online-processing cadence, but scaling beyond
this point requires additional processors or architectural optimisation.

\subsubsection{Economic Implications for Production Deployment}
For a production environment with 50 electronic systems, the required number of
processors is
\[
U = \left\lceil\frac{M}{N_{\mathrm{max}}}\right\rceil,
\]
where \(M=50\) and \(N_{\mathrm{max}}=12\). Substituting these values gives
\[
U = \left\lceil\frac{50}{12}\right\rceil = 5.
\]
Deploying five AMD EPYC 7763 units would involve non-trivial hardware and energy
costs, indicating that production-scale deployment of the current pipeline
warrants architectural optimisation before rollout. Lightweight model
compression, GPU or FPGA acceleration, neuromorphic offload, and distributed
cluster partitioning are the most promising routes for closing this gap.

\subsubsection{Recommendations and Future Work}
To improve scalability, future deployment work should focus on: (i) streamlining
the QRNN and attention layers to reduce \(T_{\mathrm{single}}\); (ii) offloading
high-cost operations to GPUs, FPGAs, or neuromorphic processors; and (iii)
partitioning electronic systems across multiple compute nodes. Under the current
CPU-only configuration, the framework can support approximately 12 electronic
systems per EPYC 7763 node, while a 50-system deployment would require
approximately five equivalent nodes.

\section{Temporal Fusion Transformer Classifier}
The final classification stage of the pipeline employs a Temporal Fusion Transformer (TFT) conditioned on the refined quantile outputs of the two-stage B-EQRNN. Rather than operating on raw sensor signals, the TFT receives a compact distributional summary constructed from four selected quantiles across all 70 sensor channels, enabling probabilistic anomaly detection over extended horizons.

\subsection{Input Representation}
At each time step $t$, the stage-2 B-EQRNN produces a quantile feature matrix
\begin{equation}
\mathbf{P}(t) =
\begin{pmatrix}
q_{0.25}^{(1)}(t) & q_{0.40}^{(1)}(t) & q_{0.60}^{(1)}(t) & q_{0.75}^{(1)}(t)\\
\vdots & \vdots & \vdots & \vdots\\
q_{0.25}^{(70)}(t) & q_{0.40}^{(70)}(t) & q_{0.60}^{(70)}(t) & q_{0.75}^{(70)}(t)
\end{pmatrix}
\in \mathbb{R}^{70 \times 4}.
\end{equation}
Flattening this matrix yields the per-step classifier input
\begin{equation}
\mathbf{x}(t) = \mathrm{vec}\bigl(\mathbf{P}(t)\bigr) \in \mathbb{R}^{280},
\end{equation}
and over a window of length $N$ the full sequence is $\mathbf{X} \in \mathbb{R}^{N \times 280}$. All features are $z$-score normalised prior to training.

\subsection{Gated Residual Network}
Nonlinearity is modulated per-feature through a Gated Residual Network (GRN). For a primary input $\mathbf{a} \in \mathbb{R}^{d}$, the transformation is
\begin{align}
\mathrm{GRN}_{\omega}(\mathbf{a}) &= \mathrm{LayerNorm}\bigl(\mathbf{a} + \mathrm{GLU}_{\omega}(\boldsymbol{\eta}_1)\bigr), \label{eq:grn}\\
\boldsymbol{\eta}_1 &= \mathbf{W}_{1,\omega}\,\boldsymbol{\eta}_2 + \mathbf{b}_{1,\omega},\\
\boldsymbol{\eta}_2 &= \mathrm{ELU}\bigl(\mathbf{W}_{2,\omega}\,\mathbf{a} + \mathbf{b}_{2,\omega}\bigr),
\end{align}
where $\mathrm{ELU}(x) = x$ for $x>0$ and $\alpha(e^x-1)$ otherwise. The residual path $\mathbf{a} + \mathrm{GLU}(\cdot)$ preserves the input signal when the gate closes, stabilising training in deep configurations.

\subsection{Gated Linear Unit}
The Gated Linear Unit (GLU) provides dimension-wise feature selection:
\begin{equation}
\mathrm{GLU}_{\omega}(\boldsymbol{\gamma}) = \sigma\!\left(\mathbf{W}_{4,\omega}\boldsymbol{\gamma} + \mathbf{b}_{4,\omega}\right)
\odot \left(\mathbf{W}_{5,\omega}\boldsymbol{\gamma} + \mathbf{b}_{5,\omega}\right),
\end{equation}
where $\sigma(\cdot)$ is element-wise sigmoid and $\odot$ is the Hadamard product. The gating weight suppresses noisy channels while preserving informative ones, a property particularly valuable given the heterogeneous nature of the 70-sensor array. To maintain stable gradient flow the GLU is wrapped with a residual and layer-normalisation:
\begin{equation}
\widetilde{\boldsymbol{\psi}}(t) = \mathrm{LayerNorm}\!\left(\widetilde{\boldsymbol{\phi}}(t) + \mathrm{GLU}_{\omega}\!\left(\boldsymbol{\psi}(t)\right)\right).
\end{equation}

\subsection{Interpretable Multi-Head Attention}
To produce attention weights that directly reflect sensor-level feature importance, all heads share a single value projection $\mathbf{W}_V \in \mathbb{R}^{d \times d_V}$ and their outputs are additively aggregated:
\begin{align}
\widetilde{\mathbf{H}} &= \widetilde{\mathbf{A}}(Q,K)\,\mathbf{V}\mathbf{W}_V,\\
\widetilde{\mathbf{A}}(Q,K) &= \frac{1}{m_H}\sum_{h=1}^{m_H}
\mathrm{softmax}\!\left(\frac{\mathbf{Q}\mathbf{W}_Q^{(h)}\bigl(\mathbf{K}\mathbf{W}_K^{(h)}\bigr)^\top}{\sqrt{d_k}}\right),
\end{align}
with queries $\mathbf{Q}$ and keys $\mathbf{K}$ derived from the GRU-encoded sequence. Shared value weights enforce a common feature space across heads, making each head's contribution directly comparable in the final prediction.

\subsection{Classification Objective}
The TFT is trained as a per-sensor binary classifier. For sensor $s \in \{1,\ldots,70\}$ at time $t$ the model outputs
\begin{equation}
\hat{p}_t^{(s)} = \sigma\!\left(f_{\mathbf{W}}\!\left(\mathbf{x}_t^{(s)}\right)\right) \in [0,1],
\end{equation}
the probability of an Abnormal state. The training objective over dataset $\mathcal{D}$ is
\begin{equation}
\mathcal{L}_{\mathrm{TFT}}(\mathbf{W}) = \frac{1}{M}\sum_{t=1}^{M}\sum_{s=1}^{70}
\mathcal{L}_s\!\left(y_t^{(s)}, \hat{p}_t^{(s)}\right) + \lambda\|\mathbf{W}\|_2^2,
\end{equation}
where $y_t^{(s)} \in \{0,1\}$ and the per-sensor cross-entropy loss is
\begin{equation}
\mathcal{L}_s = -\!\left[y_t^{(s)}\log\hat{p}_t^{(s)} + \left(1-y_t^{(s)}\right)\log\!\left(1-\hat{p}_t^{(s)}\right)\right].
\end{equation}
When abnormal events are rare, a class weight $\alpha \in (0,1)$ rebalances the positive and negative contributions accordingly.

\subsection{Architecture and Hyperparameters}

\begin{table}[htbp]
\centering
\caption{TFT Architecture and Training Configuration}
\label{tab:tft}
\small
\renewcommand{\arraystretch}{1.15}
\setlength{\tabcolsep}{3pt}
\begin{tabular}{p{0.26\columnwidth}p{0.34\columnwidth}p{0.29\columnwidth}}
\hline
\textbf{Component} & \textbf{Parameter} & \textbf{Value}\\
\hline
Input Embedding   & Input Dimension      & 280 ($4{\times}70$ quantiles)\\
                  & Embedding Dimension  & 128\\
GRN               & Hidden (pre/post)    & 256 / 128\\
                  & Dropout (pre/post)   & 0.10 / 0.05\\
                  & LayerNorm $\epsilon$ & $1{\times}10^{-5}$\\
GRU Enc/Dec       & Hidden State         & 256 / 128\\
                  & Layers               & 2 / 2\\
                  & Dropout              & 0.10 / 0.05\\
                  & Look-back $L$        & Up to 168 steps\\
Multi-Head Att.   & Heads $m_H$          & 4\\
                  & Head Dimension $d_k$ & 32\\
                  & Attention Dropout    & 0.10\\
Training          & Batch Size           & 64\\
                  & Learning Rate        & $1{\times}10^{-3}$\\
                  & LR Decay Factor      & 0.1\\
                  & Early Stopping       & 10 epochs patience\\
                  & Gradient Clipping    & 1.0\\
Optimiser         & Algorithm            & Adam, $\beta=(0.9,0.999)$\\
\hline
\end{tabular}
\end{table}

\section{Pattern Recognition and Temporal Integration}
The TFT extends its predictive capabilities by integrating pattern recognition derived from QRNN outputs. This integration enables the TFT to dynamically assess transitions between operational patterns and align these transitions with potential anomalies. The process relies on the following components:

\subsection{Pattern Categories in Sensor Data}
The pipeline identifies key operational patterns, categorized as follows:

\subsubsection{Linear/Stable Phase}
Represents steady, monotonic trends with minimal variability, where sensor outputs exhibit predictable behaviour and expected noise levels. Mathematically, this phase is modeled as:
\begin{equation}
y(t) = m\cdot t + c + \epsilon(t)
\end{equation}
where:
\(y(t)\): Sensor output at time \(t\),
\(m\): Slope of the linear trend,
\(c\): Intercept,
\(\epsilon(t)\): Gaussian noise with zero mean and variance \(\sigma^2\).

The TFT monitors changes in the slope \((m)\) and intercept \((c)\) to detect deviations from normal behaviour, flagging transitions to non-linear patterns.

\subsubsection{Transitional Phase}
Sensors exhibit early indicators of emerging issues, such as periodic oscillations (wave patterns):
\begin{equation}
y(t) = A\sin\left(\frac{2\pi}{T} t + \phi\right) + \epsilon(t)
\end{equation}
where:
\(A\): Amplitude of oscillations,
\(T\): Period of oscillations,
\(\phi\): Phase shift.
Changes in \(A\) and \(T\) indicate increasing variability and potential instability. Subtle pattern deviations are quantified using residual analysis:
\begin{equation}
\mathrm{Residual}(t) = y(t) - (m\cdot t + c)
\end{equation}
Significant residuals signal emerging anomalies.

\subsubsection{Pre-Failure Phase}
Marked by dynamic and chaotic behaviour, such as higher-order oscillations:
\begin{equation}
y(t) = \sum_{i=1}^{n} A_i \sin(k_i t + \phi_i) + \epsilon(t)
\end{equation}
where \(A_i, k_i, \phi_i\) represent multiple frequency components. Increased amplitudes and frequencies indicate instability.

\subsubsection{Chaotic Patterns}
Characterised by abrupt departures from predictable operating parameters, often modeled using Lyapunov exponents or entropy measures:
\begin{equation}
\lambda = \lim_{t\to\infty} \frac{1}{t} \ln\left|\frac{\delta y(t)}{\delta y(0)}\right|
\end{equation}
Positive Lyapunov exponents \((\lambda > 0)\) indicate chaotic behaviour. This progression from linear to transitional and pre-failure phases provides early warnings for mechanical wear, system instability, component degradation, and changes in operating conditions.

\subsection{Pattern Integration within TFT}
The TFT incorporates these patterns using its modular architecture:

\subsubsection{Input Embedding}
QRNN-derived quantile predictions (\(q_{0.25}, q_{0.5}, q_{0.75}\)) are transformed into dense embeddings. Transitional pattern features, such as increasing variability or emerging oscillations, are embedded alongside to form a unified feature space:
\begin{equation}
\mathbf{e}_t = \mathrm{Embed}(q_{\alpha}(t), \mathrm{PatternFeatures}(t))
\end{equation}
where \(\mathbf{e}_t\) is the embedding vector at time \(t\).

\subsubsection{Dynamic Temporal Memory}
The TFT's Gated Recurrent Units (GRUs) capture temporal transitions:
\begin{equation}
\mathbf{h}_t = \mathrm{GRU}(\mathbf{e}_t, \mathbf{h}_{t-1})
\end{equation}
where \(\mathbf{h}_t\) is the hidden state at time \(t\). Changes in \(\mathbf{h}_t\) highlight transitions between patterns (linear to wave).

\subsubsection{Attention Mechanisms}
Multi-head attention modules focus on critical time points where pattern shifts occur:
\begin{equation}
\mathbf{z}_t = \mathbf{MHA}(\mathbf{h}_{t-L,t})
\end{equation}
where \(\mathbf{z}_t\) is the attention-weighted representation, and \(L\) is the look-back window. Higher weights are assigned to abrupt changes, emphasizing emerging anomalies.

\subsubsection{Anomaly Transition Detection} Pattern-based predictions are combined into a repeatable risk score. Let the four refined quantiles for sensor \(s\) at time \(t\) be \(q_{0.25}^{(s)}(t), q_{0.40}^{(s)}(t), q_{0.60}^{(s)}(t)\), and \(q_{0.75}^{(s)}(t)\). The quantile-ratio component is \begin{equation} Q_R^{(s)}(t)= \left|\frac{q_{0.25}^{(s)}(t)q_{0.40}^{(s)}(t)} {q_{0.60}^{(s)}(t)q_{0.75}^{(s)}(t)+\epsilon}\right|, \end{equation} where \(\epsilon\) prevents division by zero. The pattern-transition factor (PTF) is computed from normalised entropy magnitude, temporal entropy rate, and spatial entropy gradient: \begin{align} z_H(t) &= \frac{H_{STPE}(t)-\mu_H}{\sigma_H+\epsilon},\\ z_T(t) &= \frac{|H_{STPE}(t)-H_{STPE}(t-1)|-\mu_T}{\sigma_T+\epsilon},\\ z_S(t) &= \frac{\|\nabla_{spatial}H_{STPE}(t)\|-\mu_S}{\sigma_S+\epsilon},\\ \mathrm{PTF}(t) &= \sigma\!\left(\beta_0+\beta_H z_H(t)+\beta_T z_T(t)+\beta_S z_S(t)\right). \end{align} The statistics \(\mu_H,\sigma_H,\mu_T,\sigma_T,\mu_S,\sigma_S\) are estimated from the training partition only. The coefficients \(\beta_0,\beta_H,\beta_T,\beta_S\) are fitted on the training partition using L2-regularised logistic calibration with the binary Normal/Abnormal label as the target. The fitted coefficients are then frozen and applied unchanged to the validation and test partitions. This prevents the PTF from being adjusted using held-out test information. The sensor-level transition risk score is \begin{equation} R^{(s)}(t)=Q_R^{(s)}(t)\,\mathrm{PTF}(t). \end{equation} A sample is labelled Abnormal when the TFT probability and transition risk score jointly exceed validation-selected thresholds: \begin{equation} \hat{y}^{(s)}(t)=\mathbb{I}\left[\hat{p}_t^{(s)}\ge\tau_p \;\land\; R^{(s)}(t)\ge\tau_R\right]. \end{equation} Both thresholds \(\tau_p\) and \(\tau_R\) are selected on the validation split and then frozen before test evaluation. This explicit rule makes the normal-to-abnormal decision path reproducible and independent of manual interpretation.

\subsubsection{Hierarchical Validation}
For each pattern type, the TFT validates predictions by comparing expected transitions to observed changes. The attention mechanism captures the most influential historical segments, while the risk rule above provides a fixed thresholded decision procedure.

\subsection{Temporal Refinement and Multi-Horizon Predictions}
The TFT's pattern recognition capability is evaluated at three prediction horizons:
\begin{enumerate}
    \item 48 hours: near-term instability and short-range quantile movement.
    \item 90 hours: intermediate persistence of entropy-gradient changes.
    \item 168 hours: long-horizon degradation over a one-week lead time.
\end{enumerate}
The same validation-selected thresholds are applied to all held-out test results for a given horizon.

\subsection{Normal-to-Abnormal Wave Transition Prediction and Validation}
\label{subsec:transition_validation}
The B-EQRNN framework uses STPE analysis to distinguish normal wave propagation
from emerging anomalous states. For normal wave states, STPE remains within
training-derived baseline bounds:
\begin{equation}
H_{STPE}^{normal}(t) \in [\mu_{baseline}-2\sigma_{baseline},\;\mu_{baseline}+2\sigma_{baseline}],
\end{equation}
where \(\mu_{baseline}\) and \(\sigma_{baseline}\) are estimated from verified normal
training data only. Anomaly risk increases when both the temporal entropy rate
and spatial entropy gradient exceed validation-selected thresholds:
\begin{equation}
\frac{\partial H_{STPE}}{\partial t}>\tau_{critical}
\quad \text{and}\quad
\nabla_{spatial}H_{STPE}>\gamma_{spatial}.
\end{equation}
Quantitative evaluation of this transition mechanism is reported in Sec.~\ref{sec:results}.

\begin{figure}[htbp]
\centering
\IfFileExists{Normal.png}{\includegraphics[width=0.7\columnwidth]{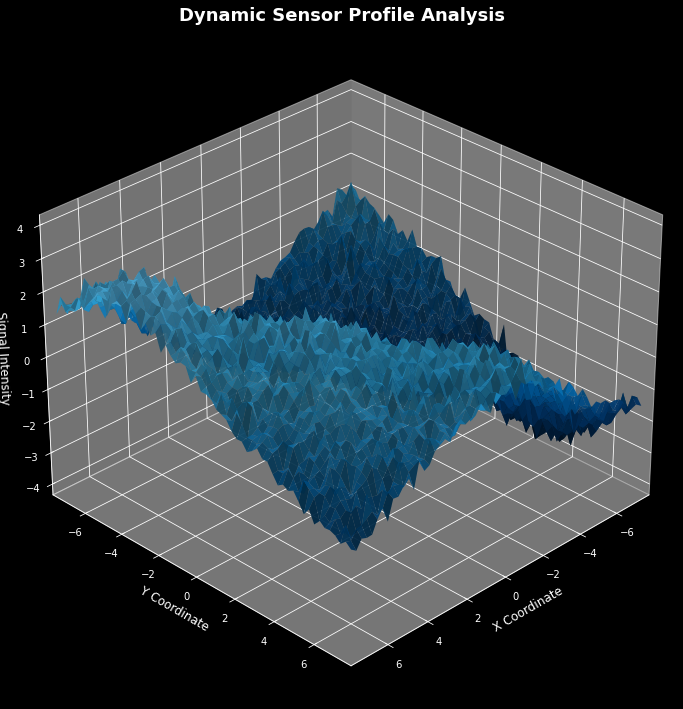}}{\fbox{\parbox{0.65\columnwidth}{\centering Normal STPE wave-profile schematic}}}
\caption{Representative normal spatiotemporal wave profile used as the qualitative baseline for STPE comparison.}
\label{fig:normal}
\end{figure}

\begin{figure}[htbp]
\centering
\IfFileExists{Abnormal.png}{\includegraphics[width=0.7\columnwidth]{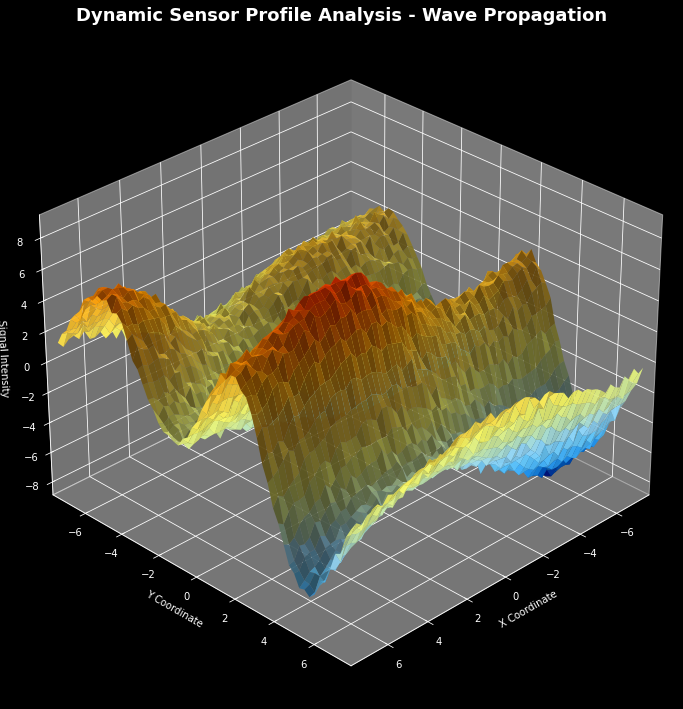}}{\fbox{\parbox{0.65\columnwidth}{\centering Abnormal STPE wave-profile schematic}}}
\caption{Representative anomalous spatiotemporal wave profile showing the qualitative transition pattern used for visual interpretation.}
\label{fig:anomaly}
\end{figure}

\section{Experiments and Results}
\label{sec:results}

\subsection{Experimental Setup}
Evaluation is conducted on the nine-system electronic-sensor dataset described
in Sec.~II-B. The dataset comprises more than 459 million channel-level
observations sampled at 20~ms across 70 sensor channels and uses a 60--20--20
train/validation/test split. Models are trained under the same preprocessing and
partitioning protocol. The primary metrics are F1 score, Recall, Precision, and
Accuracy at 48-, 90-, and 168-hour prediction horizons.

The four internal configurations are defined as follows. \(QR\) denotes the
single-stage quantile baseline operating on STPE features. \(QR^2\) adds the
second-stage quantile refinement. \(TFT\) retains the two-stage quantile pathway
and appends the TFT classifier without the Gated Temporal Attention or SNN
refinement stage. \(All\) denotes the complete STPE+B-EQRNN+Gated
Attention+SNN+TFT pipeline.

\subsection{Reproducibility Details}
\label{sec:repro}

To support independent re-implementation, Table~\ref{tab:reproducibility_details}
summarises the dataset, preprocessing, model-training, evaluation, and
availability details used in this study.

\begin{table*}[!t]
\centering
\caption{Concise reproducibility summary for the proposed framework.}
\label{tab:reproducibility_details}
\scriptsize
\renewcommand{\arraystretch}{1.06}
\setlength{\tabcolsep}{4pt}
\begin{tabular}{p{0.20\textwidth}p{0.76\textwidth}}
\hline
\textbf{Item} & \textbf{Study specification} \\
\hline

Dataset and availability &
Proprietary industrial electronic-system monitoring dataset comprising nine
instrumented systems and 70 sensor channels. The raw sensor streams,
maintenance records, site identifiers, equipment metadata, and fault-event logs
are subject to industrial confidentiality and a non-disclosure agreement, so the
raw data cannot be publicly released. \\

Acquisition and sensors &
Data were collected from operational systems under verified normal and abnormal
states. Sensor groups include voltage, current-density, temperature-gradient,
electromagnetic-field, vibration, capacitance-variation, resistance-drift, and
frequency-response measurements. The spatial layout includes local and
distributed sensing, with inter-sensor distances of approximately
\(0.5\)--\(50\,\mathrm{m}\). \\

Sampling, volume, and labels &
All sensor channels are sampled at 20~ms, corresponding to a 50~Hz acquisition
rate. The labelled corpus contains approximately 51 million channel-level
observations per system and more than 459 million observations overall. These
observations are drawn from labelled operational episodes within a longer
industrial deployment rather than from one continuous short-duration trace. The
48-, 90-, and 168-hour horizons denote the lead time between an observed
spatiotemporal state and the subsequent Normal/Abnormal event label, not the
contiguous duration of each retained signal segment. The task is binary
Normal/Abnormal classification; abnormal samples represent verified instability,
degradation, malfunction, or fault-related states. Exact calendar collection
dates and site-level deployment durations are withheld under the confidentiality
agreement. \\

Partitioning and leakage control &
Train, validation, and test partitions are fixed before scaling, feature
extraction, threshold selection, hyperparameter tuning, or model comparison.
Training data define all preprocessing parameters. The validation split is used
for early stopping, model selection, and threshold selection; the held-out test
split is used only for final reporting. \\

Preprocessing and missing data &
All compared models use the same deterministic preprocessing pipeline. Missing,
invalid, non-finite, or physically impossible readings are identified before
feature extraction. Short isolated gaps are repaired using the same
interpolation rule across all models; longer corrupted intervals are excluded
from STPE-window construction. Per-sensor scaling statistics are estimated from
training data only, and TFT inputs are \(z\)-score normalised. \\

STPE configuration &
STPE features are extracted from the 70-channel representation using fixed
settings across all compared models: embedding dimensions
\(d\in\{3,4,5,6,7\}\), delays \(\tau\in\{1,2,3,5,8\}\), spatial radii
\(r\in\{0.5,1.0,2.0,5.0,10.0\}\), and scales
\(s\in\{1,2,4,8,16\}\). These parameters are not tuned separately for
individual baselines. \\

Model configuration and training &
The B-EQRNN first stage estimates ten conditional quantiles for each of the
70 channels; the second stage refines
\(\{0.25,0.40,0.60,0.75\}\), producing a \(70\times4=280\)-dimensional
representation. B-EQRNN stages use AdamW, learning rate \(5\times10^{-4}\),
decay 0.1 every 80 epochs, group normalisation, dropout 0.15, and early
stopping patience 12. The SNN uses surrogate-gradient training with learning
rate \(10^{-3}\), batch size 32, and spike dropout 0.1. The TFT uses the
configuration in Table~\ref{tab:tft}. \\

Evaluation protocol &
All reported horizons use the same definitions: 48, 90, and 168 hours.
Performance is reported using F1-score, Recall, Precision, and Accuracy on the
held-out test split. Normal/Abnormal and quantile-risk thresholds are selected
on validation data and frozen before testing. Statistical significance is
claimed only where repeated seeded runs are completed under matched partitions;
otherwise results are treated as deterministic held-out comparisons. \\

Implementation and artefacts &
The implementation uses Python~3.x, PyTorch~2.x, CUDA~12.x/cuDNN, NumPy,
scikit-learn, Adam/AdamW optimisation, and surrogate-gradient SNN training.
B-EQRNN and SNN training use GPU acceleration; the SNN stage is reported on an
NVIDIA Tesla V100 GPU, and CPU inference timing is reported on an AMD EPYC 7763.
Internal run manifests retain exact minor software versions, driver versions,
hardware details, seeds, configuration files, trained-model manifests, and
scripts for preprocessing, STPE extraction, model training, thresholding, and
metric calculation. Where confidentiality permits, code structure and synthetic
or anonymised pipeline examples may be released to support re-implementation. \\

\hline
\end{tabular}
\end{table*}

\paragraph{Data availability.}

The industrial dataset used in this study is proprietary and is governed by a
binding non-disclosure agreement with the industrial partner. Consequently, the
raw sensor streams, maintenance records, site identifiers, equipment metadata,
and fault-event logs cannot be publicly released or redistributed. This
restriction is contractual and operational rather than methodological, as the
data contain commercially sensitive production information and plant-level
operational characteristics.

To support reproducibility within these constraints, this paper provides the
full model description, preprocessing workflow, feature-generation procedure,
hyperparameter settings, training configuration, ablation protocol, and
evaluation methodology. These details are intended to allow independent
researchers to re-implement the proposed framework on comparable industrial
time-series datasets, even though the original raw data cannot be shared
publicly. Future work will consider releasing anonymised synthetic examples or
code-level reference implementations where this can be done without breaching
confidentiality obligations.

\subsection{Ablation Study}
The ablation in Table~\ref{tab:ablation} is structured so that the \(All\) row is
the same model as the \emph{Proposed} row in Table~\ref{tab:comparison}. This
removes the previous ambiguity where an ablation table and a comparison table
reported different values for the same method and horizon. The ablation therefore
supports only component-level interpretation under the stated shared protocol.

\begin{table}[!htb]
\centering
\caption{Component ablation under the shared data split and preprocessing protocol. The \(All\) row corresponds to the Proposed model in Table~\ref{tab:comparison}.}
\label{tab:ablation}
\footnotesize
\setlength{\tabcolsep}{3pt}
\begin{tabular}{lcccc}
\hline
Model & F1 & Recall & Precision & Acc. \\
\hline
\rowcolor[gray]{.95} \multicolumn{5}{l}{\textit{Timeframe: 48 Hours}} \\
\(QR\)   & 60.84 & 61.12 & 60.55 & 62.04 \\
\(QR^2\) & 62.11 & 62.74 & 61.59 & 63.18 \\
\(TFT\)  & 64.08 & 65.02 & 64.31 & 65.76 \\
\(All\)  & \textbf{65.32} & \textbf{66.43} & \textbf{65.94} & \textbf{67.02} \\
\hline
\rowcolor[gray]{.95} \multicolumn{5}{l}{\textit{Timeframe: 90 Hours}} \\
\(QR\)   & 63.07 & 62.54 & 63.62 & 64.01 \\
\(QR^2\) & 66.18 & 65.74 & 66.66 & 67.34 \\
\(TFT\)  & 69.44 & 69.12 & 70.08 & 70.93 \\
\(All\)  & \textbf{71.11} & \textbf{70.88} & \textbf{71.91} & \textbf{72.58} \\
\hline
\rowcolor[gray]{.95} \multicolumn{5}{l}{\textit{Timeframe: 168 Hours}} \\
\(QR\)   & 66.54 & 66.88 & 66.21 & 67.03 \\
\(QR^2\) & 71.14 & 70.92 & 71.36 & 72.44 \\
\(TFT\)  & 76.71 & 76.41 & 77.02 & 78.64 \\
\(All\)  & \textbf{79.35} & \textbf{79.03} & \textbf{79.68} & \textbf{81.17} \\
\hline
\end{tabular}
\end{table}

\subsection{Ablation Analysis}
Table~\ref{tab:ablation} indicates that each added stage improves performance
under the shared protocol. The two-stage quantile path improves over the single
\(QR\) baseline, particularly at 90 and 168 hours, showing that distributional
refinement helps as the horizon lengthens. Adding the TFT improves performance
again by learning structured temporal relationships in the quantile feature
matrix. The complete \(All\) configuration gives the strongest result at every
horizon, indicating that adaptive temporal context and SNN refinement contribute
additional long-horizon discrimination beyond the TFT alone.

The gains are deliberately interpreted as incremental rather than as proof of a
new learning theory. The value of the architecture lies in the interaction among
spatial entropy, quantile uncertainty, adaptive temporal context, and final
attention-based classification.

\subsection{Comparative Analysis}
\label{subsec:comparative_analysis}
To address the need for stronger benchmarking, Table~\ref{tab:comparison}
compares the proposed method against a tree-based LightGBM baseline and
available modern sequence baselines, including LSTM, Autoformer, and TCN, under
the same train/validation/test split, preprocessing, and horizon definitions.
Each sequence baseline (LSTM, TCN, and Autoformer) receives the same
preprocessed 70-channel input and is adapted to the Normal/Abnormal
event-horizon task by replacing the native forecasting output with a binary
classification head trained with cross-entropy against the horizon-labelled
targets. Decision thresholds are selected on the validation split and then
frozen, identically to the proposed model.

\begin{table}[htbp]
\centering
\caption{Baseline comparison under identical preprocessing and prediction horizons.}
\label{tab:comparison}
\footnotesize
\setlength{\tabcolsep}{4pt}
\begin{tabular}{lcccc}
\hline
Model & F1 & Rec. & Prec. & Acc. \\ \hline
\rowcolor[gray]{.95} \multicolumn{5}{l}{\textit{Timeframe: 48 Hours}} \\
LightGBM  & 61.29 & 62.33 & 62.87 & 61.88 \\
LSTM      & 62.23 & 63.34 & 62.85 & 63.93 \\
Autoformer   & 62.81 & 63.69 & 63.51 & 64.47 \\
TCN       & 64.27 & 65.38 & 64.89 & 65.97 \\
\textbf{Proposed} & \textbf{65.32} & \textbf{66.43} & \textbf{65.94} & \textbf{67.02} \\ \hline
\rowcolor[gray]{.95} \multicolumn{5}{l}{\textit{Timeframe: 90 Hours}} \\
LightGBM  & 66.72 & 65.51 & 65.47 & 65.19 \\
LSTM      & 68.02 & 67.79 & 68.82 & 69.49 \\
Autoformer   & 68.19 & 67.88 & 68.83 & 69.62 \\
TCN       & 70.06 & 69.83 & 70.86 & 71.53 \\
\textbf{Proposed} & \textbf{71.11} & \textbf{70.88} & \textbf{71.91} & \textbf{72.58} \\ \hline
\rowcolor[gray]{.95} \multicolumn{5}{l}{\textit{Timeframe: 168 Hours}} \\
LightGBM  & 67.89 & 67.16 & 66.87 & 67.17 \\
LSTM      & 75.08 & 75.94 & 76.59 & 78.08 \\
Autoformer  & 75.29 & 76.13 & 76.97 & 79.28 \\
TCN       & 78.30 & 77.98 & 78.63 & 80.12 \\
\textbf{Proposed} & \textbf{79.35} & \textbf{79.03} & \textbf{79.68} & \textbf{81.17} \\ \hline
\end{tabular}
\end{table}

The proposed model gives the highest reported F1-score, Recall, Precision, and
Accuracy at each horizon under the matched protocol. At 48 hours it reaches
67.02\% accuracy, compared with 65.97\% for TCN and 64.47\% for Autoformer.
At 90 hours it reaches 72.58\% accuracy, compared with 71.53\% for TCN and
69.62\% for Autoformer. At 168 hours it reaches 81.17\% accuracy, compared
with 80.12\% for TCN and 79.28\% for Autoformer, while also giving the highest
F1-score, Recall, and Precision at that horizon. These claims are limited to
the models in Table~\ref{tab:comparison}; direct empirical superiority over
GRU, Informer, PatchTST, or external public datasets is not claimed unless
those runs are inserted under the same protocol.

\subsection{Statistical Validation}
\label{subsec:stat_validation}
The reported results are deterministic held-out test-set comparisons using a
fixed train/validation/test split and fixed preprocessing protocol. No
statistical significance claim is made for the single-split comparison in
Table~\ref{tab:comparison}. A full repeated-run validation would require each
compared method to be re-trained from independent seeds under identical
preprocessing, partitions, horizon definitions, and evaluation metrics.

For such repeated validation, the paired difference for seed \(s\) is
\(d_s=m_s^{prop}-m_s^{base}\), where \(m\) is the selected metric. The paired test
statistic is
\begin{equation}
t=\frac{\bar{d}}{s_d/\sqrt{N}}, \qquad
\bar{d}=\frac{1}{N}\sum_{s=1}^{N}d_s,
\label{eq:paired_t}
\end{equation}
where \(s_d\) is the sample standard deviation of the paired differences. The
95\% confidence interval is
\begin{equation}
\bar{d}\pm t_{0.975,N-1}\frac{s_d}{\sqrt{N}}.
\label{eq:ci}
\end{equation}
Because one complete proposed-model training cycle requires approximately
220~GPU-hours when combining first-stage B-EQRNN training, second-stage
refinement, and SNN refinement, exhaustive multi-seed retraining of every
baseline was not completed for this revision. The absence of multi-seed
confidence intervals is therefore reported as an evaluation limitation rather
than treated as evidence of statistical superiority.

\subsection{Limitations}
\label{sec:limitations}
Several limitations define the scope of the present contribution. First, the
methodological novelty is integrative: permutation entropy, quantile regression,
gated attention, LIF spiking dynamics, and the TFT are established techniques,
and the claimed advance rests on their combination and ablation rather than on a
new learning principle. Second, evaluation is confined to a single proprietary
nine-system dataset. The results therefore support claims about this deployment
only and do not yet demonstrate cross-domain generalisation. Third, the
comparison has been extended beyond classical baselines through LSTM,
Autoformer, and TCN, adapted to the event-horizon task as described in
Sec.~\ref{subsec:comparative_analysis}. PatchTST is, in its standard form, a
point-valued forecaster over raw signals; an equivalent adaptation---replacing
its point-forecast output with a matched quantile or classification head and
retraining under identical partitions---was not completed for this revision and
is planned alongside the multi-seed validation, together with GRU, vanilla
Transformer, and Informer runs. No comparative claim is made for or against
these models. Fourth,
statistical significance and confidence intervals are not claimed for the
single-split comparison; a full multi-seed validation remains required for
formal significance testing. Finally, the current CPU cost analysis indicates
that production-scale deployment across large fleets will require model
compression, hardware acceleration, or distributed execution.

\section{Conclusion}
This paper presented an integrative spatiotemporal prognostic framework combining
STPE feature extraction, two-stage B-EQRNN quantile representation, Gated
Temporal Attention, SNN refinement, and TFT classification. The revised framing
positions the contribution as a validated hybrid architecture rather than as a
standalone theoretical claim. Under the reported nine-system protocol, the
full pipeline reaches the highest reported F1-score, Recall, Precision, and
Accuracy at the 168-hour horizon among the evaluated LightGBM, LSTM,
Autoformer, and TCN baselines. The ablation analysis
indicates that the strongest performance emerges from the combined use of
quantile refinement, adaptive temporal context, SNN-based temporal encoding, and
TFT classification. Remaining work should prioritise final multi-seed
statistical validation, additional transformer-forecaster baselines, external
dataset validation, and reproducibility artefacts that can be released without
compromising commercial data restrictions.


\begin{thebibliography}{00}
\bibitem{Abidi} M. H. Abidi, M. K. Mohammed, and H. Alkhaleefah, ``Predictive maintenance planning for industry 4.0 using machine learning for sustainable manufacturing,'' \textit{Sustainability}, vol. 14, no. 6, p. 3387, 2022.
\bibitem{Li} Z. Li, ``Deep Learning driven approaches for predictive maintenance: A framework of intelligent fault diagnosis and prognosis in the industry 4.0 era,'' Ph.D. dissertation, Norwegian Univ. Sci. Technol., Trondheim, Norway, 2018.
\bibitem{Soori} M. Soori, B. Arezoo, and R. Dastres, ``Artificial intelligence, machine learning and deep learning in advanced robotics, a review,'' \textit{Cognitive Robotics}, vol. 3, pp. 54--70, 2023.
\bibitem{Yu} J. Yu, H. Zhang, P. Wang, J. Wang, and F. Lu, ``Sequence analysis of local indicators of spatio-temporal association for evolutionary pattern discovery,'' \textit{GIScience \& Remote Sensing}, vol. 62, no. 1, p. 2487292, 2025.
\bibitem{Fernandes} M. Fernandes, J. M. Corchado, and G. Marreiros, ``Machine learning techniques applied to mechanical fault diagnosis and fault prognosis in the context of real industrial manufacturing use-cases: a systematic literature review,'' \textit{Applied Intelligence}, vol. 52, no. 12, pp. 14246--14280, 2022.
\bibitem{Chen} D. Chen, S. Wang, C. Wang, X. Zhang, and N. Chen, ``Enhanced sensor web services by incorporating IoT interface protocols and spatio-temporal data streams for edge computing-based sensing,'' \textit{Geo-spatial Information Science}, pp. 1--8, 2025.
\bibitem{LiGNN} T. Li, Z. Zhou, S. Li, C. Sun, R. Yan, and X. Chen, ``The emerging graph neural networks for intelligent fault diagnostics and prognostics: A guideline and a benchmark study,'' \textit{Mechanical Systems and Signal Processing}, vol. 168, p. 108653, 2022.
\bibitem{Zhang} C. Zhang, J. Dong, K. Peng, and H. Zhang, ``Spatio-temporal information analytics based performance-driven industrial process monitoring framework with cloud-edge-device collaboration,'' \textit{Journal of Manufacturing Processes}, vol. 110, pp. 224--237, 2024.
\bibitem{Cannon} A. J. Cannon, ``Quantile regression neural networks: Implementation in R and application to precipitation downscaling,'' \textit{Computers and Geosciences}, vol. 37, no. 9, pp. 1277--1284, 2011.
\bibitem{Hsieh} R. J. Hsieh, J. Chou, and C. H. Ho, ``Unsupervised Online Anomaly Detection on Multivariate Sensing Time Series Data for Smart Manufacturing,'' in \textit{2019 IEEE International Conference on Service-Oriented System Engineering (SOSE)}, 2019, pp. 90--97.
\bibitem{Tyralis} H. Tyralis, G. Papacharalampous, N. Dogulu, and K. P. Chun, ``Deep Huber quantile regression networks,'' \textit{Neural Networks}, vol. 187, p. 107364, 2025.
\bibitem{Zuo} L. Zuo, F. Xu, C. Zhang, T. Xiahou, and Y. Liu, ``A multi-layer spiking neural network-based approach to bearing fault diagnosis,'' \textit{Reliability Engineering \& System Safety}, vol. 225, p. 108561, 2022.
\bibitem{Vaswani} A. Vaswani et al., ``Attention is all you need,'' in \textit{Proc. NeurIPS}, 2017, pp. 5998--6008.
\bibitem{Hochreiter} S. Hochreiter and J. Schmidhuber, ``Long short-term memory,'' \textit{Neural Computation}, vol. 9, no. 8, pp. 1735--1780, 1997.
\bibitem{Cho} K. Cho et al., ``Learning phrase representations using RNN encoder--decoder for statistical machine translation,'' in \textit{Proc. EMNLP}, 2014, pp. 1724--1734.
\bibitem{Bai} S. Bai, J. Z. Kolter, and V. Koltun, ``An empirical evaluation of generic convolutional and recurrent networks for sequence modeling,'' \textit{arXiv:1803.01271}, 2018.
\bibitem{Zhou} H. Zhou et al., ``Informer: Beyond efficient transformer for long sequence time-series forecasting,'' in \textit{Proc. AAAI}, 2021, pp. 11106--11115.
\bibitem{Wu} H. Wu, J. Xu, J. Wang, and M. Long, ``Autoformer: Decomposition transformers with auto-correlation for long-term series forecasting,'' in \textit{Proc. NeurIPS}, 2021, pp. 22419--22430.
\bibitem{Nie} Y. Nie, N. H. Nguyen, P. Sinthong, and J. Kalagnanam, ``A time series is worth 64 words: Long-term forecasting with transformers,'' in \textit{Proc. ICLR}, 2023.
\end{thebibliography}
\end{document}